\documentclass[3p,times,procedia]{elsarticle}
\usepackage{nupha_ecrc}

\volume{00}

\firstpage{1}

\journalname{Nuclear Physics A}

\runauth{}

\jid{nupha}

\jnltitlelogo{Nuclear Physics A}

\usepackage{amssymb}

\usepackage{lineno}

\usepackage[figuresright]{rotating}

\begin{document}

\begin{frontmatter}

\dochead{}

\title{Low-Mass Dielectron Production in pp, p--Pb and Pb--Pb Collisions with ALICE}

\author{Patrick Reichelt (for the ALICE Collaboration)}

\address{Institut f\"ur Kernphysik, Goethe-Universit\"at Frankfurt am Main, Germany}


\begin{abstract}
The ALICE Collaboration measures the production of low-mass dielectrons in pp, p--Pb and Pb--Pb collisions at the LHC. The main detectors used in the analyses are the Inner Tracking System, Time Projection Chamber and Time-Of-Flight detector, all located at mid-rapidity. The dielectron yield in p--Pb collisions shows an overall agreement with the hadronic cocktail. The pair transverse momentum distributions are sensitive to the contributions from open heavy-flavours. In Pb--Pb collisions, uncorrected background-subtracted yields have been extracted in two centrality classes. In pp collisions the production of virtual photons relative to the inclusive yield is determined by analyzing the dielectron excess with respect to the expected hadronic sources. The direct photon cross section is then calculated and found to be in agreement with NLO pQCD calculations. A feasibility study for LHC Run 3 after the ALICE upgrade indicates the possibility for a future measurement of the early effective temperature.
\end{abstract}


\end{frontmatter}


\section{Introduction}
The measurement of electron-positron pairs in the low invariant mass region allows studying the vacuum and in-medium properties of light vector mesons. Additionally, dielectrons from semileptonic decays of correlated heavy-quark mesons carry information on the heavy-flavour production in the different collision systems. Low-mass dielectrons are also produced by internal conversion of virtual direct photons. To quantify modifications of the dielectron production in heavy-ion collisions, measurements in pp collisions serve as a reference, while the analysis of p--A collisions allows disentangling cold from hot nuclear matter effects. In ALICE \cite{Ali-Exp} at the LHC, dielectron measurements are performed using the central barrel detectors at mid-rapidity. In this proceedings we present the invariant mass analyses in p--Pb and Pb--Pb collisions, a mass-differential study in p--Pb, and a virtual direct photon measurement in pp collisions. Prospects of a future measurement in Pb--Pb collisions after the ALICE upgrade for LHC Run 3 are also discussed.

\section{Low-mass dielectrons in ALICE: analysis and results}
The analyses presented here are based on $3 \cdot 10^{8}$ minimum-bias pp collisions at $ \sqrt{s} = 7 {\rm ~TeV}$, $10^{8}$ minimum-bias p--Pb collisions at $ \sqrt{s_{\rm NN}} = 5.02 {\rm ~TeV}$, as well as $1.7 \cdot 10^{7}$ and $1.2 \cdot 10^{7}$ events in central (0--10\%) and semi-central (20--50\%) Pb--Pb collisions, respectively. The electron selection in the pp and p--Pb analyses share the same fiducial cuts on transverse momentum ($p_{\rm T} > 0.2 {\rm ~GeV}/c$) and pseudorapidity ($ |\eta| < 0.8$). In Pb--Pb a tighter cut of $0.4 < p_{\rm T} < 3.5 {\rm ~GeV}/c$ ensures exclusion of charged pions and reduces the contribution of soft electrons from conversions and $\pi^0$-Dalitz decays to the combinatorial background. A conversion rejection cut on the pair level is done in all collision systems. In pp collisions, a clean electron sample is achieved by applying cuts on the time-of-flight (TOF) and on the specific energy loss ${\rm d}E/{\rm d}x$ in the Time Projection Chamber (TPC). In p--Pb and Pb--Pb, a TOF signal is not required in order to increase the electron efficiency, yet used if available to purify the electron sample. Kaons and protons are rejected by an electron inclusion cut on the ${\rm d}E/{\rm d}x$ measured by the Inner Tracking System (ITS). Dielectron spectra are created for unlike-sign (ULS) and like-sign (LS) combinations of selected particles. The signal yield is obtained as $N_{\rm ULS}^{\rm same} - N_{\rm LS}^{\rm same} \cdot R$, where $R$ accounts for any acceptance difference between unlike- and like-sign pairs due to detector effects. It is built via a mixed-event technique as $R = N_{\rm ULS}^{\rm mix} / N_{\rm LS}^{\rm mix}$ and stays within 5\% around unity for the full invariant mass range considered in these analyses.

\begin{figure}[h]
\begin{minipage}{17pc}
\vspace{0.45pc}
\includegraphics[width=16.7pc]{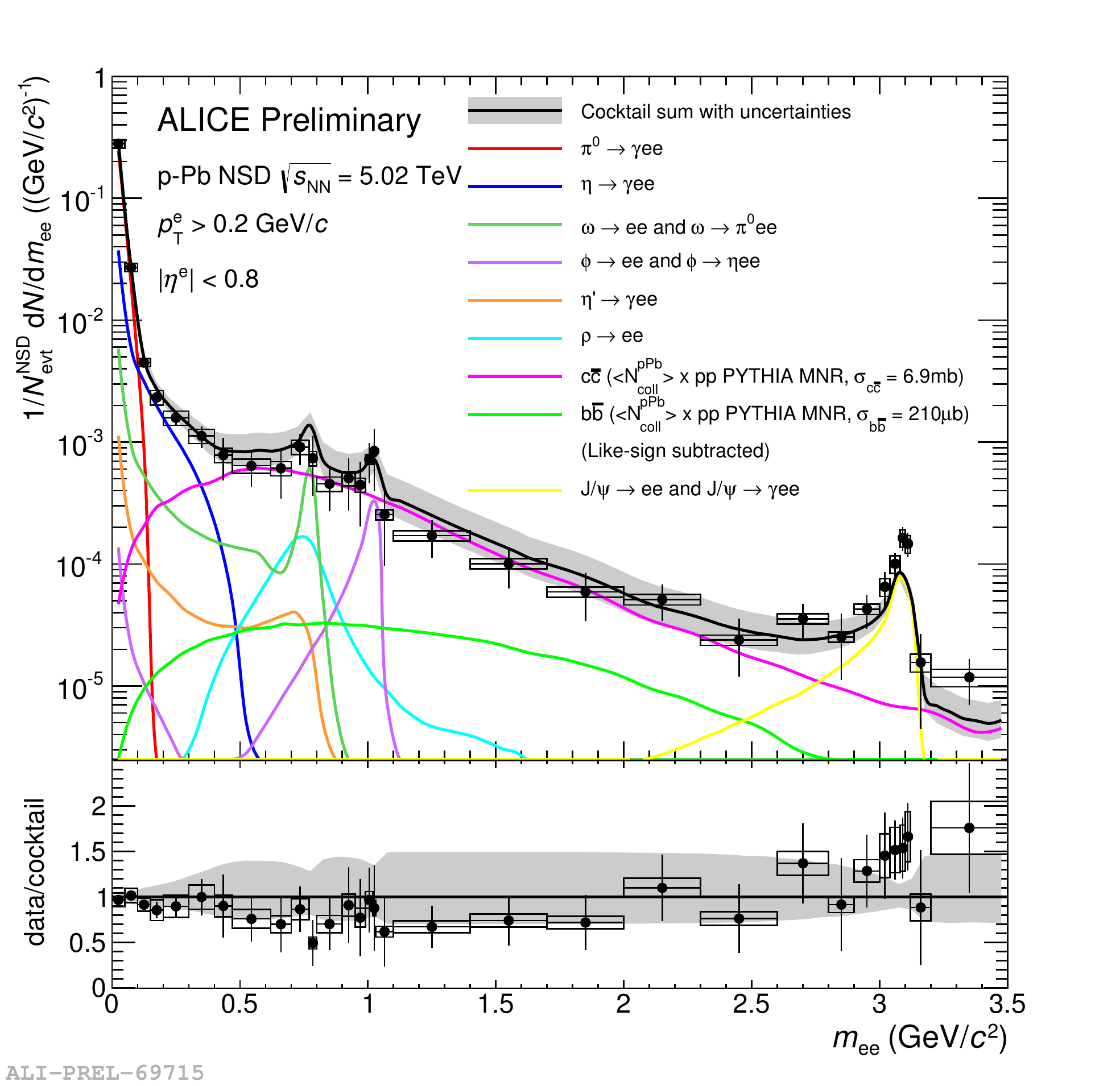}
\caption{\label{pPb-mee}Dielectron mass spectrum in p--Pb collisions in comparison to the hadronic cocktail.}
\end{minipage}\hspace{1.7pc}%
\begin{minipage}{17pc}
\includegraphics[width=16.7pc]{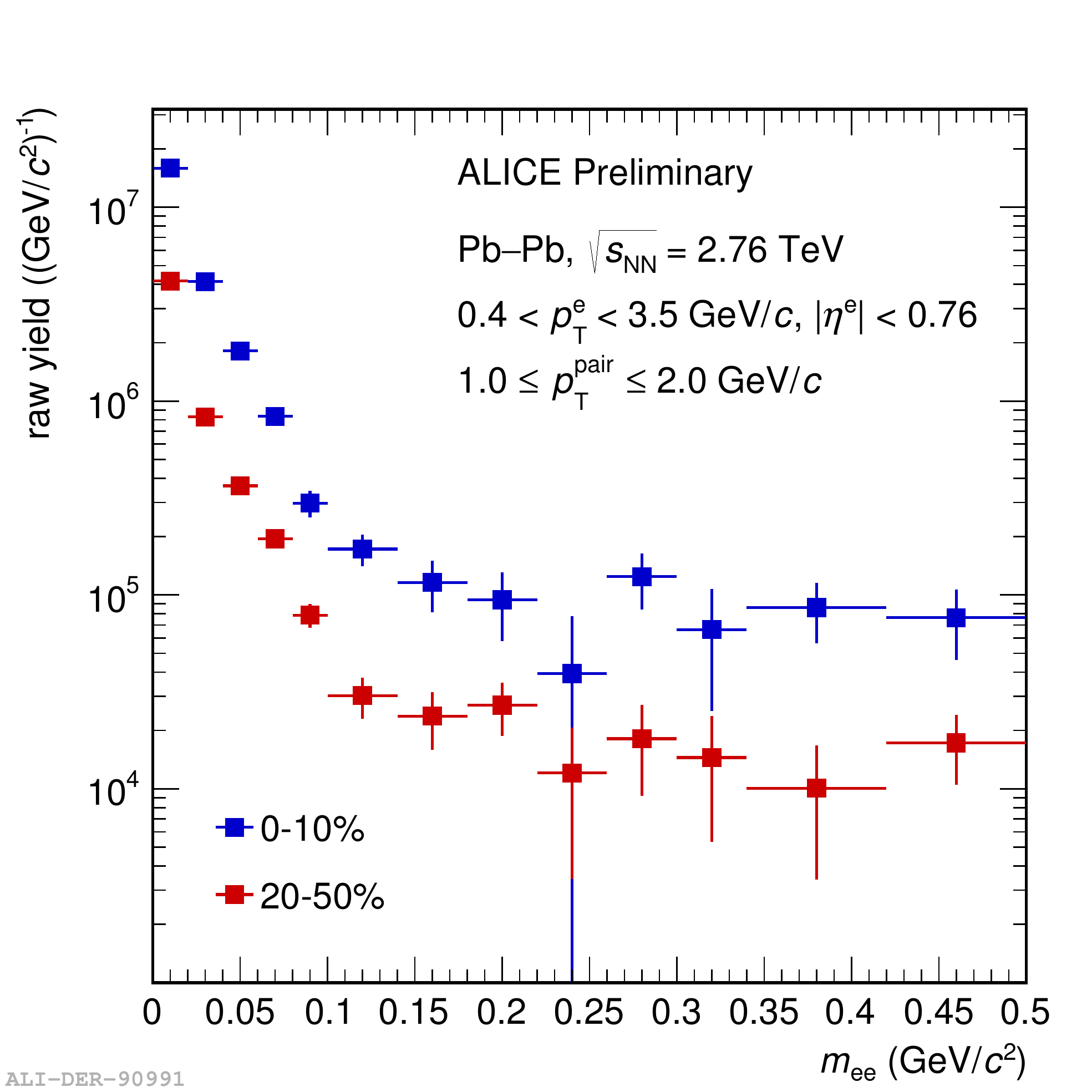}
\caption{\label{PbPb-mee}Dielectron raw subtracted yield in central and semi-central Pb--Pb collisions after background subtraction.}
\end{minipage}
\end{figure}

\begin{figure}[t]
\begin{minipage}{20pc}
\vspace{0.2pc}
\includegraphics[height=14.6pc]{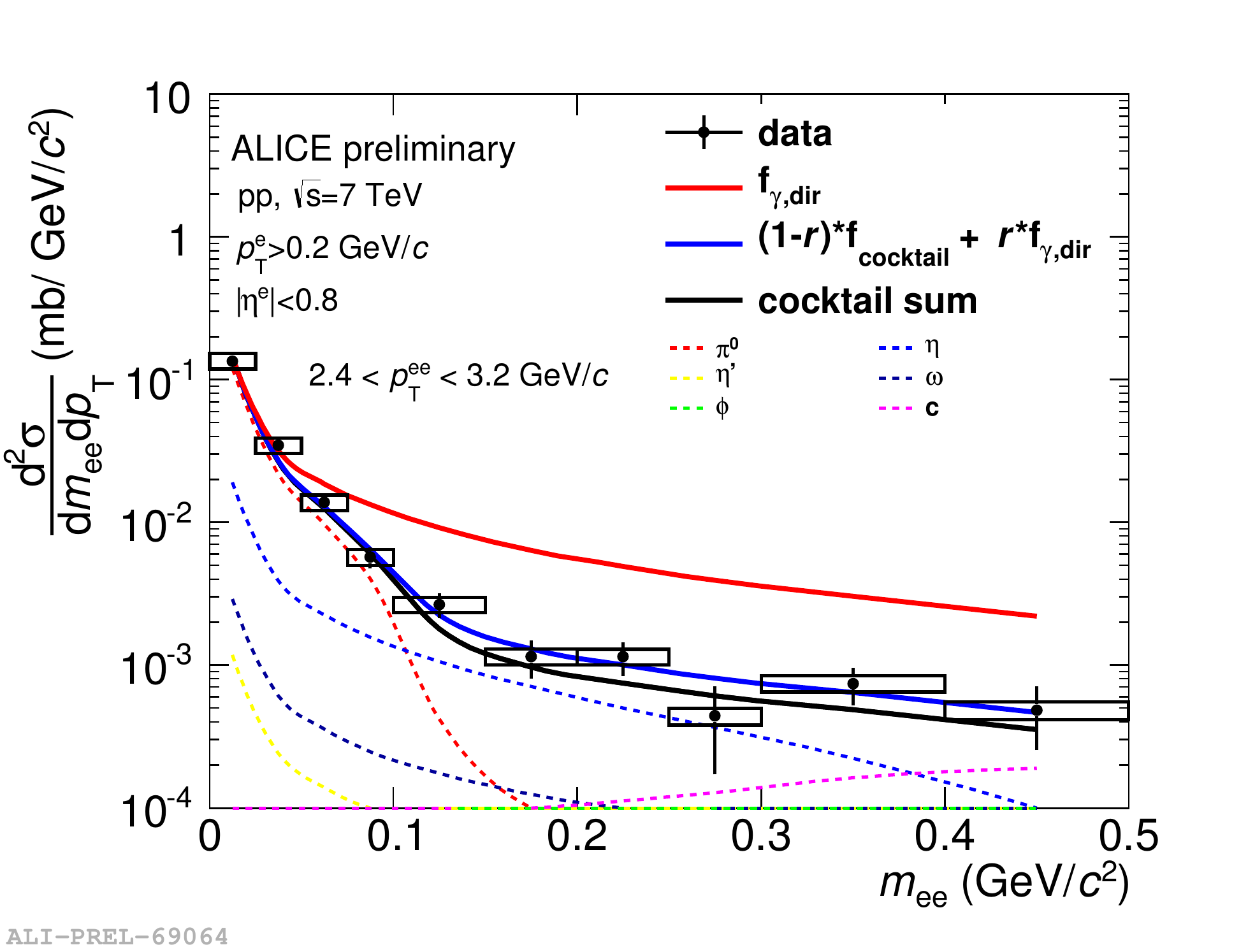}
\caption{\label{pp-fit}Low-mass region of the dielectron cross section in pp collisions for $2.4 < p_{\rm T}^{\rm ee} < 3.2 {\rm ~GeV}/c$ compared to the hadronic cocktail, as well as to a fit to extract the virtual direct photon fraction $r$.}
\end{minipage}\hspace{1.7pc}%
\begin{minipage}{14pc}
\includegraphics[height=14.6pc]{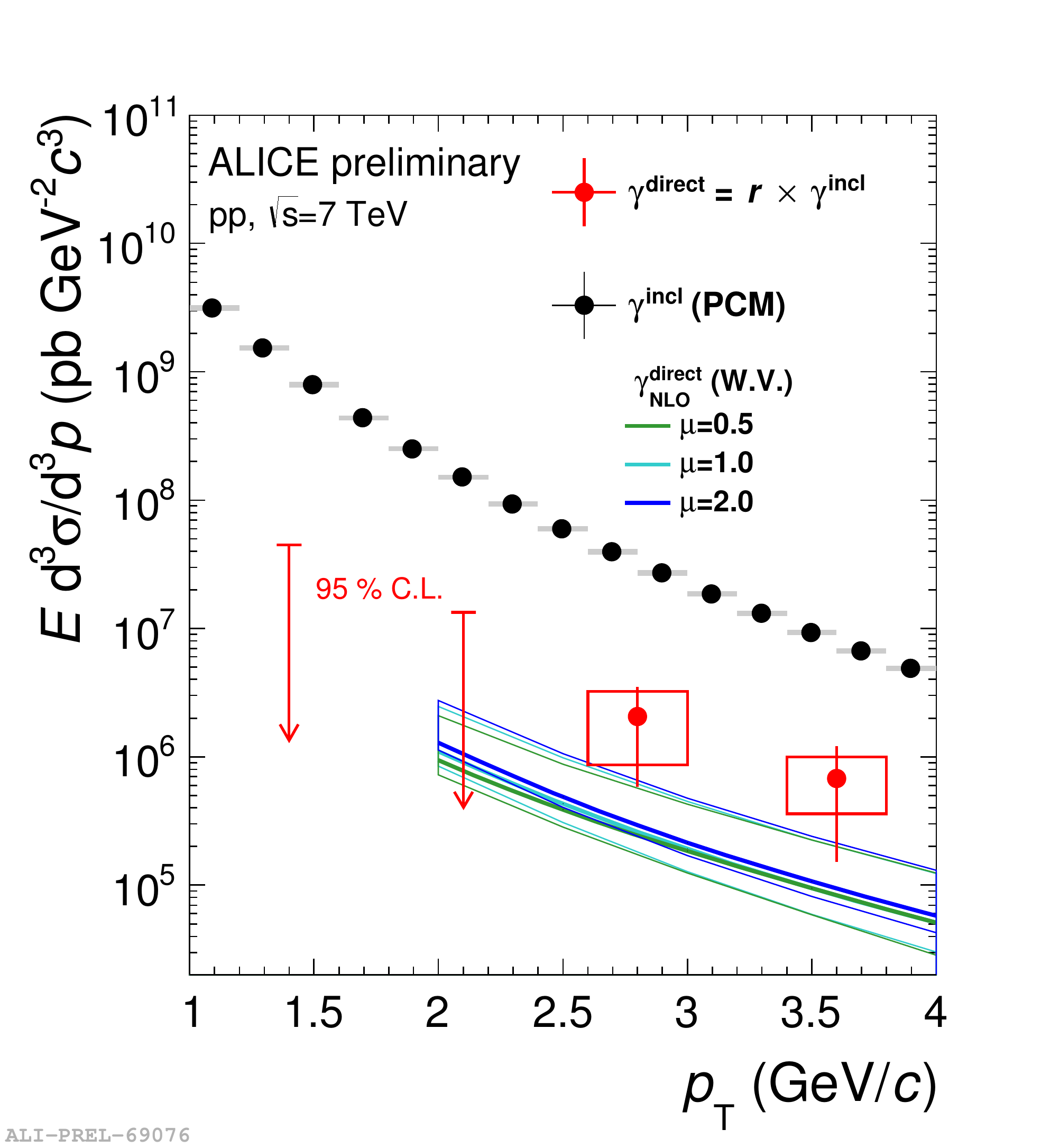}
\caption{\label{pp-photons}Direct photon cross section compared to NLO pQCD calculations, and the inclusive photon cross section extracted from PCM.}
\end{minipage}
\end{figure}

\begin{figure}[h]
\begin{minipage}{17pc}
\includegraphics[width=16.2pc]{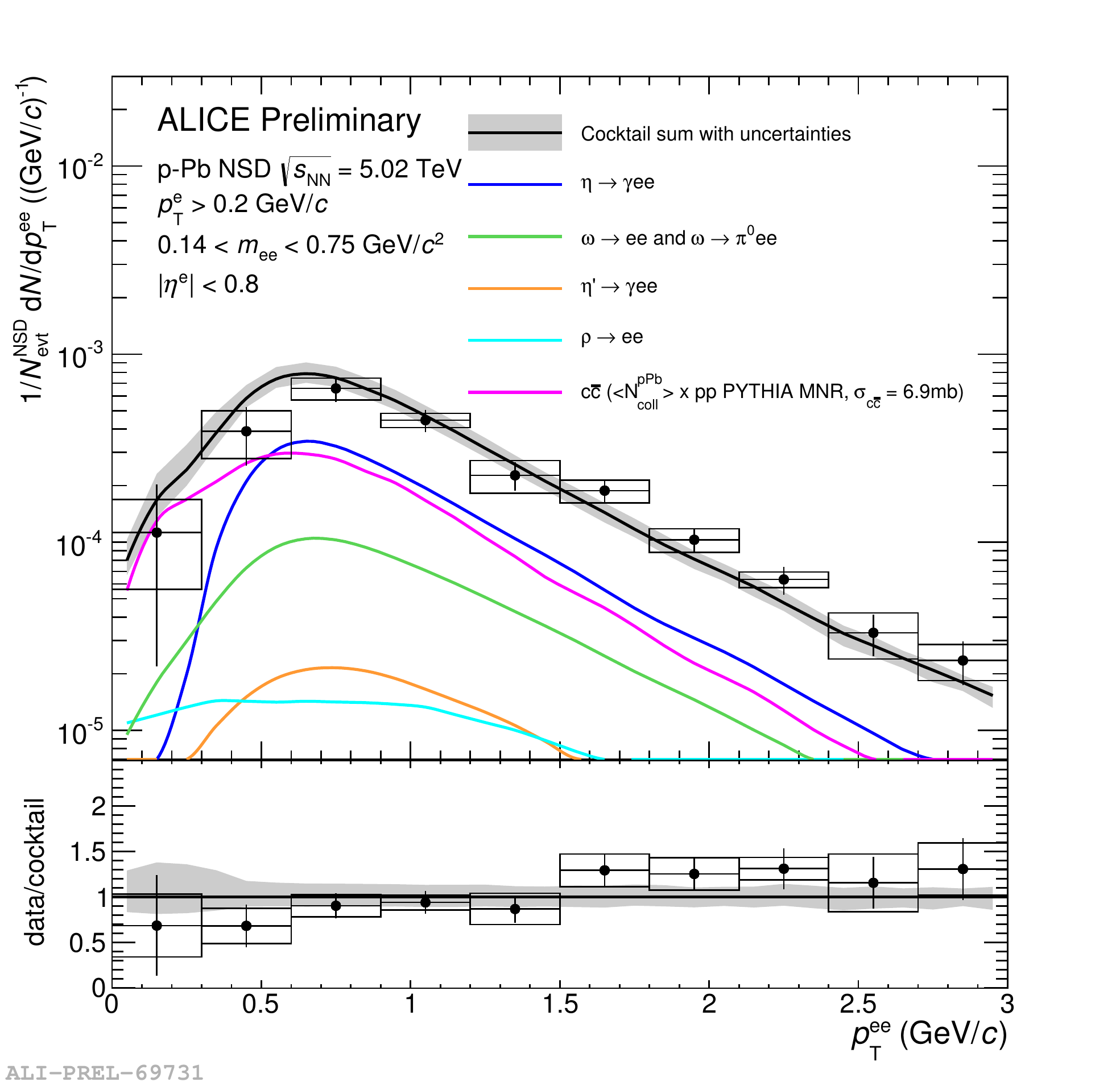}
\caption{\label{pPb-ptee-2}Dielectron pair-$p_{\rm T}$ spectrum in p--Pb collisions compared to the hadronic cocktail, for $0.14 < m_{\rm ee} < 0.75 {\rm ~GeV}/c^2$.}
\end{minipage}\hspace{1.7pc}%
\begin{minipage}{17pc}
\includegraphics[width=16.2pc]{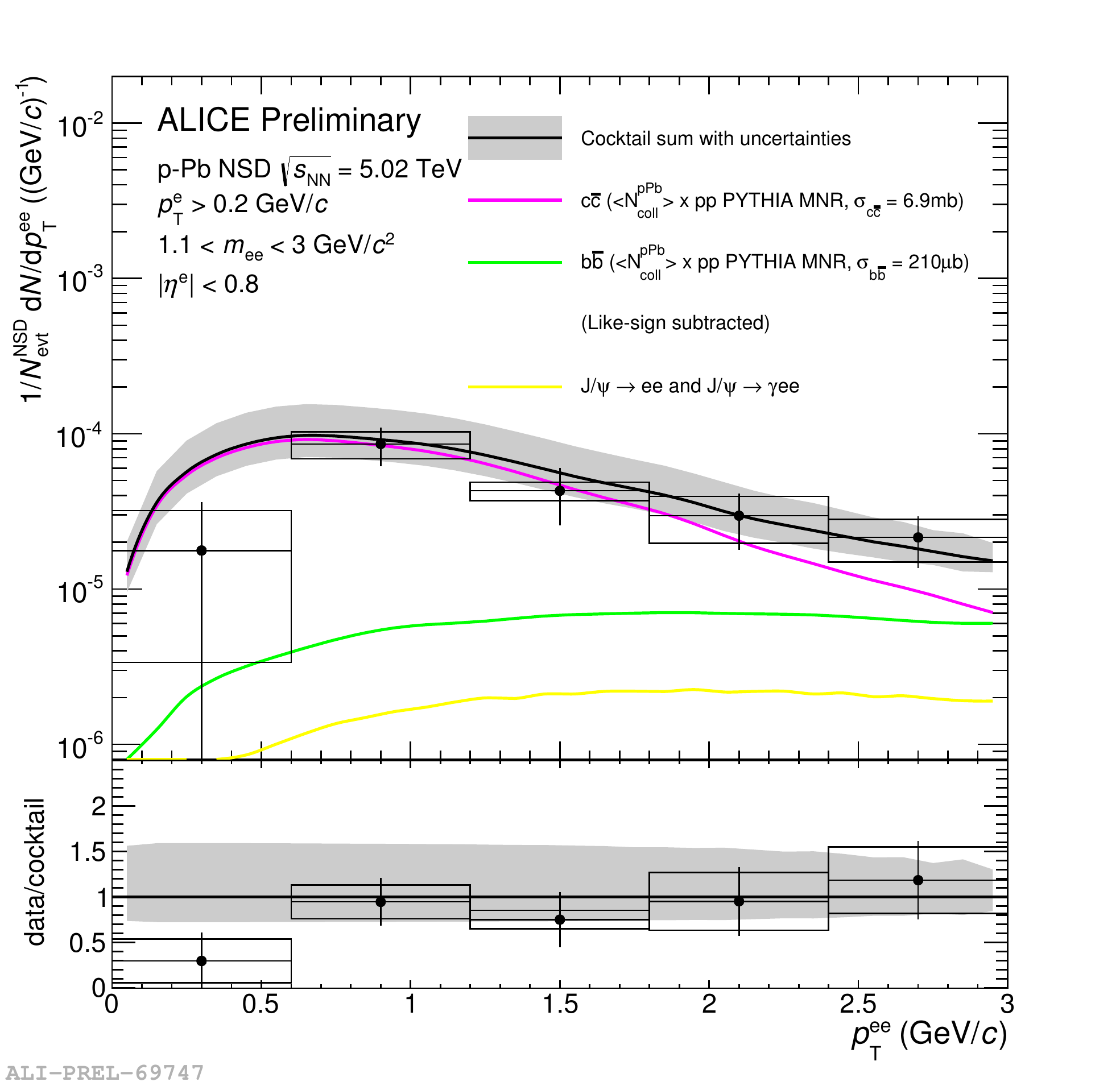}
\caption{\label{pPb-ptee-4}Same as Fig. \ref{pPb-ptee-2}, but for $1.1 < m_{\rm ee} < 3.0 {\rm ~GeV}/c^2$, which is most sensitive to $c\bar{c}$ and $b\bar{b}$ production.}
\end{minipage}
\end{figure}

The corrected dielectron yield in p--Pb collisions, integrated over pair-$p_{\rm T}$ ($p_{\rm T}^{\rm ee}$), is shown in Fig. \ref{pPb-mee}. The data points and systematic uncertainties are extracted from the mean values and the spread of results obtained with $22$ different combinations of analysis cut settings (inspired by \cite{Barlow}). The data are compared to the hadronic cocktail, which uses the charged pion measurement \cite{Ali-pi-charged} as $\pi^0$ input and $m_{\rm T}$-scaling for the other light mesons. Open heavy-flavour and $J/\psi$ contributions are calculated from PYTHIA, tuned to independent ALICE measurements in pp and p--Pb collisions \cite{Ali-HF, Ali-Jpsi}. Uncertainties coming from the input of the cocktail sources are shown as a grey band in Fig. \ref{pPb-mee}. Data and cocktail are in reasonable agreement within their uncertainties over the full mass range. Figure \ref{PbPb-mee} shows, for Pb--Pb collisions in both centralities, the background-subtracted uncorrected yields in the low-mass region for $1 < p_{\rm T}^{\rm ee} < 2 {\rm ~GeV}/c$. The corresponding signal-to-background ratios within $0.2 <  m_{\rm ee} < 0.5 {\rm ~GeV}/c^2$ are $0.01$ for 20--50\% and $0.003$ for 0--10\% centrality. The efficiency correction and cocktail comparison are in progress.\\
The kinematic region $p_{\rm T}^{\rm ee} \gg m_{\rm ee}$ is useful to study the production of virtual direct photons. Figure \ref{pp-fit} shows the differential dielectron cross-section in pp collisions at low $m_{\rm ee}$ for one pair-$p_{\rm T}$ region, compared to a hadronic cocktail and its components. Also shown is the expected mass distribution of dielectrons coming from virtual direct photons, after having applied the single-electron fiducial cuts on $p_{\rm T}$ and $\eta$. A fit to the data is performed to determine the virtual photon fraction $r$ by using the function $f_{\rm combined} = (1-r) \cdot f_{\rm cocktail} + r \cdot f_{\rm \gamma,dir}$ in the range $0.1 < m_{\rm ee} < 0.4 {\rm ~GeV}/c^2$. This fit is done in four pair-$p_{\rm T}$ regions and the extracted virtual photon fraction is then multiplied by the inclusive photon cross section, also measured by ALICE via the photon conversion method (PCM) \cite{Ali-PCM}. In Figure \ref{pp-photons} the resulting direct photon cross section is compared to NLO pQCD calculations. Reasonable agreement is found, within uncertainties, between data and model. Details on this analysis can be found in \cite{Koehler}.\\
The dielectron yield in p--Pb collisions is analyzed as a function of pair-$p_{\rm T}$ in different mass regions to gain additional sensitivity in the hadronic cocktail comparison. At LHC energies, heavy-flavour contributions already become relevant for $0.14 < m_{\rm ee} < 0.75 {\rm ~GeV}/c^2$, as shown in Fig. \ref{pPb-ptee-2}, while they completely dominate the spectrum for $1.1 < m_{\rm ee} < 3.0 {\rm ~GeV}/c^2$, as shown in Fig. \ref{pPb-ptee-4}. The analysis is being extended to higher momenta to evaluate different model calculations and extract heavy-flavour cross sections.

\section{Upgrade Study for future dielectron measurements in Pb--Pb collisions}
\begin{figure}[h]
\begin{minipage}{17pc}
\includegraphics[width=16.6pc]{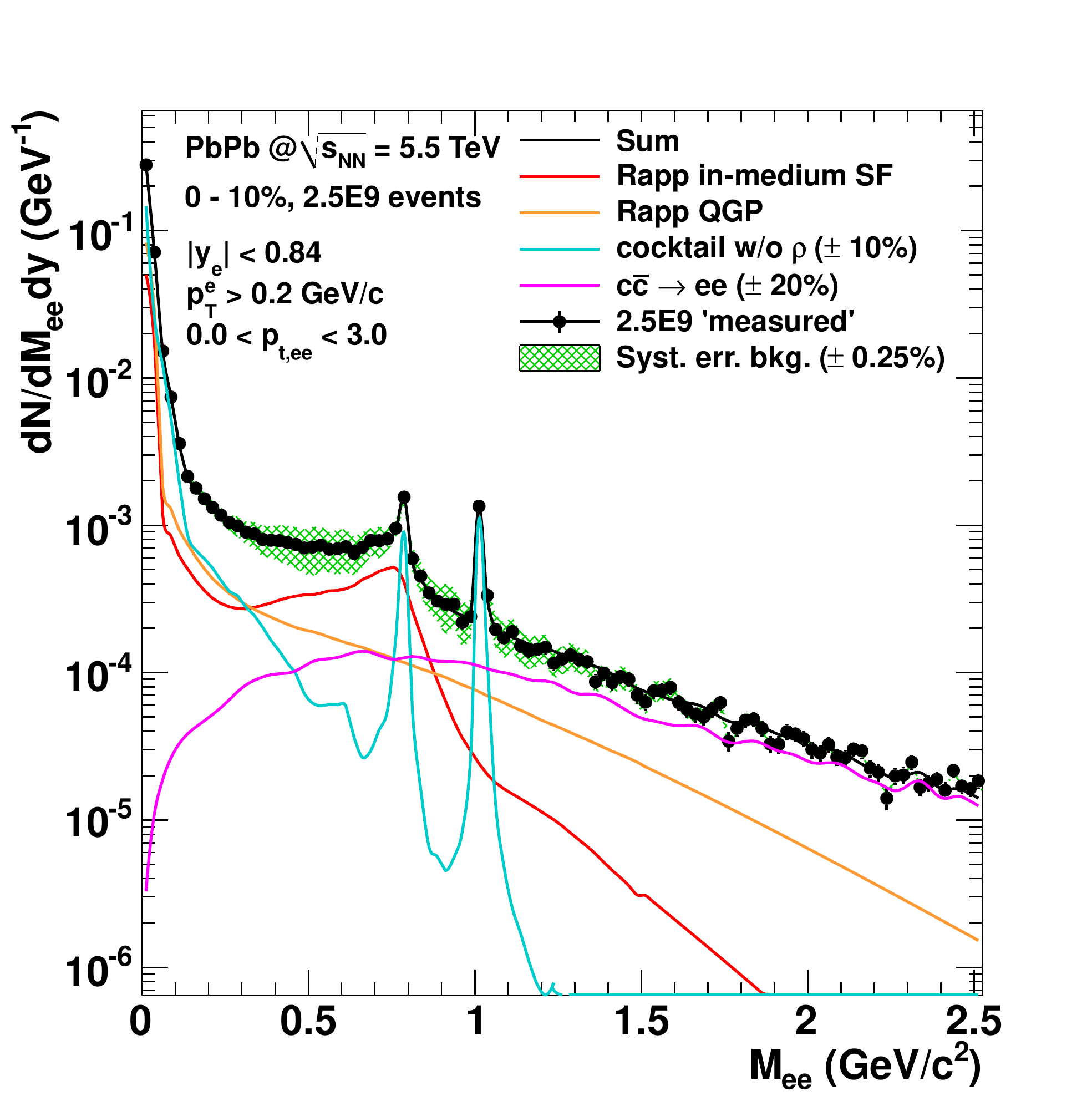}
\caption{\label{signal-upg}Predicted dielectron signal composition for $ \sqrt{s_{\rm NN}} = 5.5 {\rm ~TeV}$, with applied DCA cut, and expected precision of datapoints using the upgraded experimental setup of ALICE. \cite{Ali-Upg-ITS-TDR}}
\end{minipage}\hspace{1.7pc}%
\begin{minipage}{17pc}
\includegraphics[width=16.6pc]{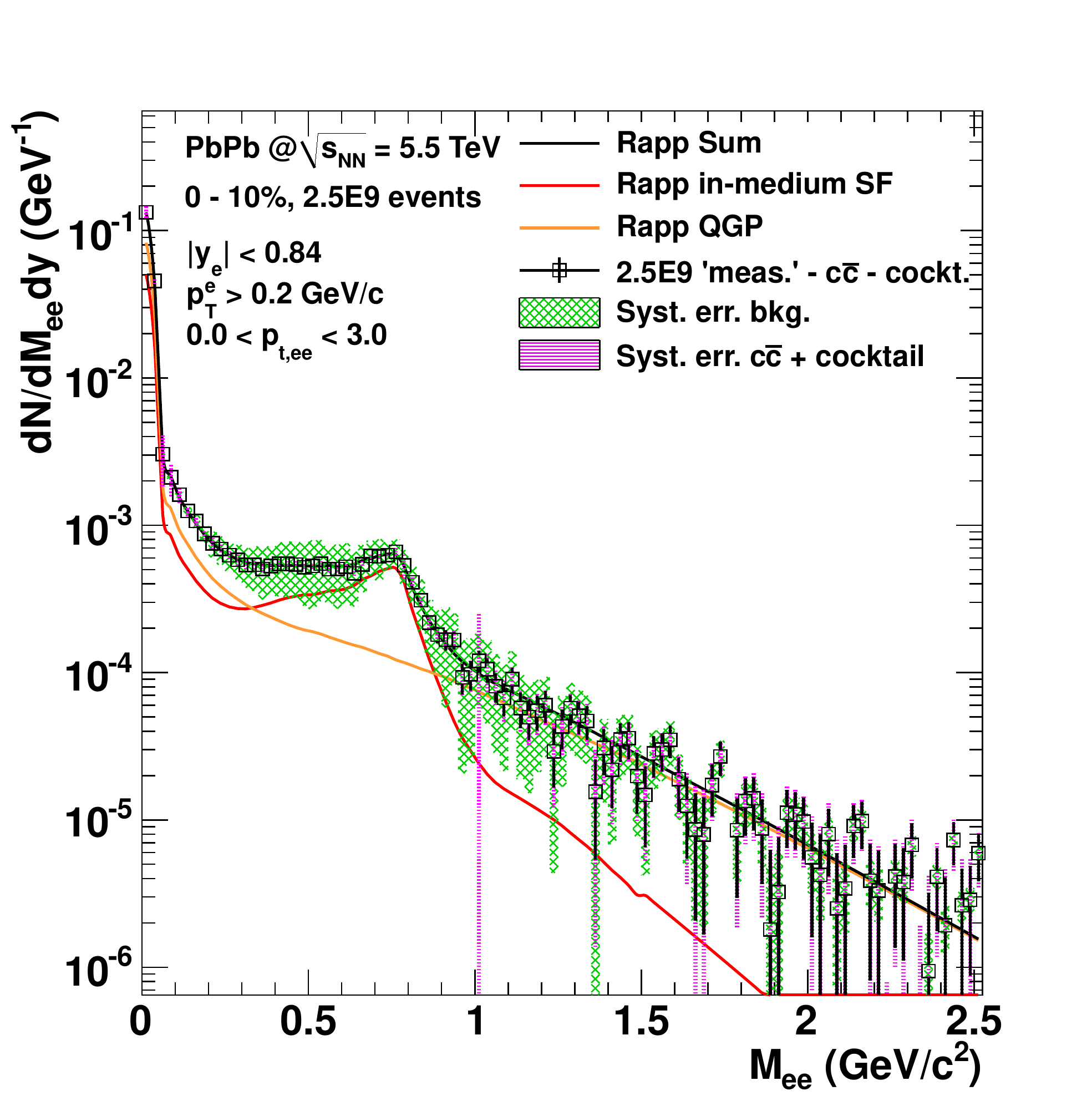}
\caption{\label{excess-upg}Excess dielectron yield after hadronic cocktail subtraction and propagating expected uncertainties. An exponential fit for $m_{\rm ee} > 1.1 {\rm ~GeV}/c^2$ gives the early effective temperature. \cite{Ali-Upg-ITS-TDR}}
\end{minipage}
\end{figure}

A fine-binning differential low-mass dielectron measurement in Pb--Pb collisions is a major physics case for the ALICE upgrade for LHC Run 3 \cite{Ali-Upg-LOI}. We evaluated to which precision a possible thermal excess yield \cite{Rapp-1999} could be reconstructed in Pb--Pb collisions at full LHC energy ($ \sqrt{s_{\rm NN}}=$ 5.5 TeV). A sample of electron candidates is created according to the expected detector performance after the upgrade, additionally being operated at a reduced solenoid magnetic field to increase the acceptance. A statistical significance is calculated from realistically constructed signal and background spectra, and is then used for Poisson sampling of data points around the model input, both shown in Fig. \ref{signal-upg}. A DCA cut (distance of closest approach between track and vertex) is applied to suppress the $c\bar{c}$ contribution. Figure \ref{excess-upg} shows the expected spectrum after subtraction of the hadronic sources, where a thermal fit can be applied. The ALICE upgrade is expected to allow for a temperature measurement in the inter\-mediate mass region within $10\%$ statistical (based on $2.5 \cdot 10^{9}$ events) and $20\%$ systematic uncertainty \cite{Ali-Upg-ITS-TDR}.

\section{Conclusions}
The dielectron yield measured in p--Pb collisions is in agreement with the expected hadronic sources, and the $p_{\rm T}^{\rm ee}$-distributions are sensitive to heavy-flavour production. From Pb--Pb collisions, raw subtracted yields at low mass have been extracted in two centrality classes. In the related kinematic region $p_{\rm T}^{\rm ee} \gg m_{\rm ee}$, virtual direct photons are observed in pp collisions. The extracted direct photon cross section agrees with NLO pQCD calculations within uncertainties. A feasibility study indicates that the ALICE upgrade for LHC Run 3 will facilitate a measurement of the effective temperature at the early times of a Pb--Pb collision.

\paragraph{Acknowledgements}
This work is supported by the German BMBF and the Helmholtz Association.





\end{document}